# A large starphene comprising pentacene branches

*Jan Holec, Beatrice Cogliati, James Lawrence, Alejandro Berdonces-Layunta, Pablo Herrero, Yuuya Nagata, Marzena Banasiewicz, Boleslaw Kozankiewicz, Martina Corso, Dimas G. de Oteyza, Andrej Jancarik\* and Andre Gourdon\**

[*]    **Dr. A. Jancarik,**
       Centre d'Elaboration de Matériaux et d'Etudes Structurales, CEMES-CNRS,
       29 rue Jeanne Marvig, 31055 Toulouse, France

       Institute of Organic Chemistry and Biochemistry of Czech Academy of Science, IOCB CAS
       Flemingovo nám. 542, 160 00 Praha 6, Czech Republic
       E-mail: <andrej.jancarik@cemes.fr>

[*]    **Dr. A. Gourdon, Dr. J. Holec**
       Centre d'Elaboration de Matériaux et d'Etudes Structurales, CEMES-CNRS,
       29 rue Jeanne Marvig, 31055 Toulouse, France
       E-mail: andre.gourdon@cemes.fr

**Beatrice Cogliati**
Centre d'Elaboration de Matériaux et d'Etudes Structurales, CEMES-CNRS, 29 rue Jeanne Marvig, 31055 Toulouse, France
Current address : Dipartimento di Scienze Chimiche, della Vita e della Sostenibilità Ambientale, Università di Parma, Parco Area delle Scienze 17/A, I-43124 Parma, Italy;
**Dr. J. Lawrence, A. Berdonces-Layunta, P. Herrero, Dr. M. Corso, Dr. D. G. de Oteyza**
Donostia International Physics Center, 20018 San Sebastián, Spain; Centro de Fisica de Materiales, CSIC-UPV/EHU, 20018 San Sebastián, Spain
**Dr. Y. Nagata**
Japan Institute for Chemical Reaction Design and Discovery (WPI-ICReDD), Hokkaido University, Sapporo, Hokkaido 001-0021, Japan
**Dr. M. Banasiewicz, Dr. B. Kozankiewicz**
Institute of Physics, Polish Academy of Sciences, Al. Lotników 32/46, 02-668 Warsaw, Poland

***Abstract:*** *Starphenes are attractive compounds due to their characteristic physicochemical properties that are inherited from acenes, making them interesting compounds for organic electronics and optics. However, the instability and low solubility of larger starphene homologs make their synthesis extremely challenging. Herein, we present a new strategy leading to pristine [16]starphene in preparative scale. Our approach is based on a synthesis of a carbonyl-protected starphene precursor that is thermally converted in a solid-state form to the neat [16]starphene, which is then characterised with a variety of analytical methods, such as 13C CP-MAS NMR, TGA, MS MALDI, UV-Vis and FTIR spectroscopy. Furthermore, high-resolution STM experiments unambiguously confirm its expected structure and reveal a moderate electronic delocalisation between the pentacene arms. Nucleus-independent chemical shifts NICS(1) are also calculated to survey its aromatic character.*

**Introduction**

Starphenes are 2D polyaromatic hydrocarbons (PAHs) comprising three acene[1] arms connected through a single benzene ring.[2] These structures can be used in single-molecule electronics as NAND[3] or NOR[4] gates. Furthermore, as a sort of 2D PAHs, starphenes can represent attractive materials for applications in organic electronics, as components in organic field-effect transistors (OFETs) or organic light-emitting diodes (OLEDs).[5]

In general, the synthesis of starphenes is a challenging task due to their limited solubility, caused by strong intermolecular π-π stacking.[6] Therefore, only a few examples of unsubstituted starphenes have been reported until now and the largest one, decastarphene-(3,3,3), was described as far back as 1968 by Clar and Mullen.[7]

To overcome this solubility problem, pendant groups are often introduced at the molecular peripheries, enhancing both solubility and stability of the starphenes.[6] For example, [7]starphene derivatives with n-butyl[8] or hexyloxy groups[9] were synthesized in this manner. Recently, Bunz et al. have prepared [13]- and [16]starphene derivatives with bulky triisopropylsilyl-ethynyl groups,[10] which are the longest substituted starphenes prepared to date. However, these bulky substituents can modify the electronic properties of the molecules, their packing and moreover affect interactions with metal surfaces, which can be limiting factors in the applications of the substituted starphenes in organic and single molecule electronics. The most applied synthetic strategy towards these soluble starphenes is a palladium-catalysed [2+2+2] cycloaddition reaction of in situ generated arynes from appropriate arene derivatives.[11–13] Another often used approach is a nickel-mediated Yamamoto coupling reaction of ortho-dibrominated arenes.[2,6,10,14]

As an alternative, on-surface synthesis (OSS) in ultra-high vacuum (UHV) allows the preparation, although in minute amounts, of organic structures that are hardly attainable by traditional in-solution chemistry.[15–17] Recently, Fasel et al.[18] prepared a [13]starphene from ortho-dibromotetracene by OSS on a Ag(111) surface. Then, Grill et al.[19] reported OSS of [10]starphene from ortho-dibromoantracene controlled by the surface structure.

However, in order to apply these unsubstituted starphene molecules as semi-conducting materials in organic electronic devices, or as single molecular gates, new practical methods of their preparation are still of great interest.

Herein, we report a synthesis of planar and fully conjugated [16]starphene 1 without any attached substituents (Figure 1).

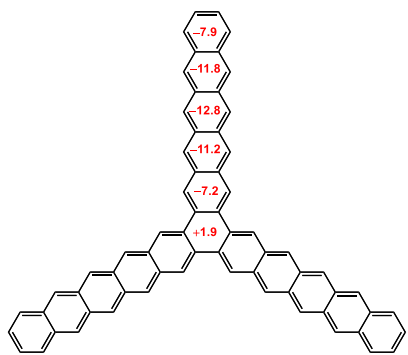

*Figure 1*. Structure of the [16]starphene 1; the red numbers represent NICS(1) values calculated at the B3LYP/6-311++g(2d,p)//B3LYP/6-31G(d,p) level of theory (vide infra).

Furthermore, we describe its comprehensive characterisation by FTIR, UV-Vis spectroscopy, 13C CP-MAS NMR, TGA and MS MALDI. In addition, high-resolution scanning tunnelling microscopy (HR-STM) experiments precisely confirm structure 1, and are complemented by scanning tunnelling spectroscopy (STS) measurements that provide an estimate of the energy gap between the molecule's highest occupied and lowest unoccupied molecular orbitals (HOMO and LUMO, respectively). To gain further insight into its electronic properties, we also carried out theoretical calculations. To the best of our knowledge, compound 1 is the largest unsubstituted starphene that has been synthesised to date.

**Results and Discussion**

Synthesis of [16]Starphene 1. We have synthesised [16]starphene 1 in 7 steps (Scheme 1), starting from the 7,7-dimethoxy-2,3,5,6-tetramethylenebicyclo[2.2.1]-heptane 2.[20,21]

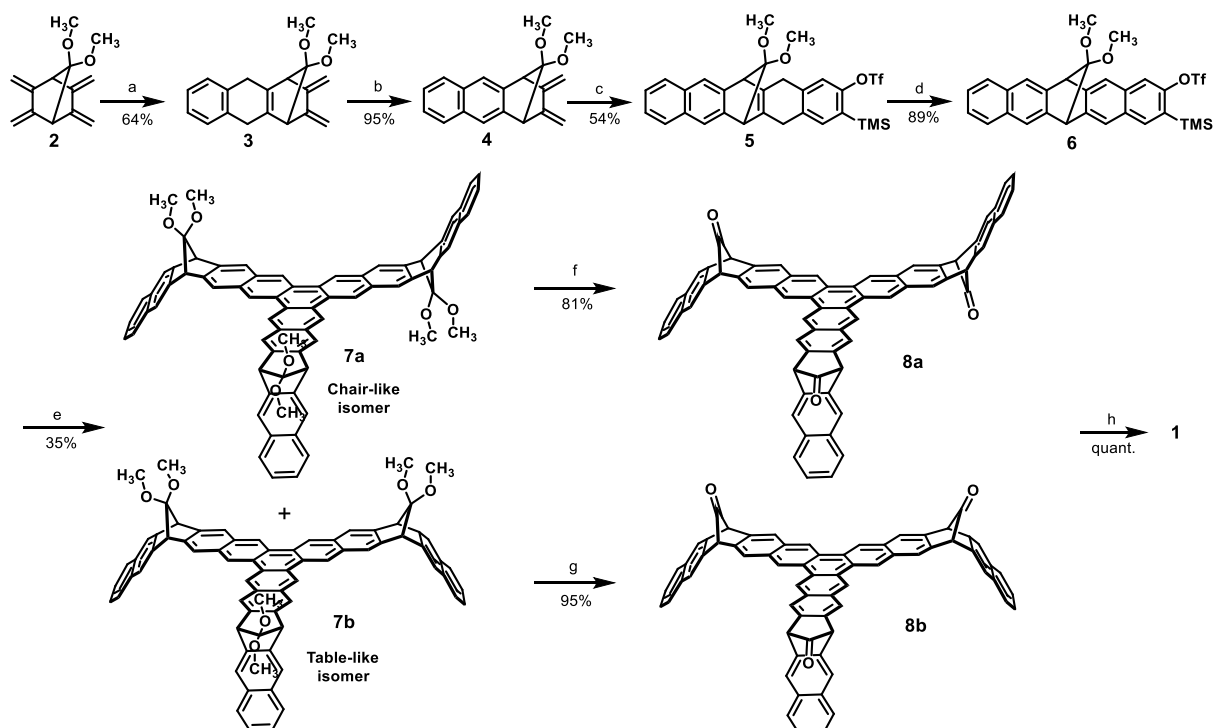

*Scheme 1.* a) 2-(Trimethylsilyl)phenyl trifluoromethanesulfonate, (1.3 eq.), CsF (2.0 eq.), $CH_3CN$, 25 °C, 16 h ; (b) DDQ (1.0 eq.), toluene, 25 °C, 3 h; (c) 1,4-bis(trimethylsilyl)phenyl-2, 5-bis(triflate) (1.2 eq.), CsF (1.2 eq.). $CH_3CN$, 25 °C, 16 h; (d) DDQ (1.5 eq.), toluene, 25 °C, 2 h; (e) CsF (3.5 eq.), $Pd_2(dba)_3$ (10 mol%), $CH_3CN$, 25 °C, 16 h; (f) TMSI (6.0 eq.), DCM, 25 °C, 18 h; (g) TMSI (6.0 eq.), DCM, 25 °C, 18 h; (h) neat, 170 °C.

First, a Diels-Alder (DA) reaction of benzyne, generated in situ from commercially available 2-(trimethylsilyl)phenyl-trifluoromethanesulfonate, with one diene side of 2, gave 3 in 64% yield. After oxidation by 2,3-dichloro-5,6-dicyano-p-benzoquinone (DDQ), diene 4 was produced in 95% yield. Under a controlled DA reaction of the monoaryne species, selectively generated from 1,4-bis(trimethylsilyl)-phenyl-2,5-bis(triflate)[22] and diene 4, compound 5 was prepared in 54% yield. Oxidation of 5 by DDQ gave 6 in 89% yield. The bridging ketal moiety in 6 ensures its solubility by preventing strong intermolecular π-π stacking, which is crucial for the construction of extended polyaromatic structures. The preserved trimethylsilyl and triflate groups in 6 are in a place for in situ formation of the aryne employed in the next synthetic step. It is worth mentioning that compound 6 is a very valuable building block for the preparation of a large variety of PAHs.[23,24] Under the optimised palladium-catalysed [2+2+2] cycloaddition reaction,[6,11] the aryne precursor 6 was converted to a mixture of two expected isomers 7a and 7b (in a ratio 2:1, see Supporting Information for a chromatogram, Figure S30) with a total yield of 35%. The subtle difference in polarity between 7a and 7b allowed us to separate these isomeric products by standard column chromatography and their distinct symmetry enabled the determination of their structures by comparing their 1H NMR spectra. The spectrum of 7a was consistent with its lower C2v symmetry (chair-like structure). Four signals (3.27, 2×3.28, and 3.29 ppm) of methyl groups of ketal moieties appeared in the spectrum – two signals for the inner methyl groups pointing to the central part of the starphene structure and then another two signals of outer methyls pointing to the ends of the arms. In the case of isomer 7b, with C3v point symmetry (table-like structure), only two of those signals (3.27 and 3.29 ppm) were observed (See Supporting Information, Figures S11 and S13). The last reaction step in solution was the cleavage of three ketal groups by trimethylsilyl iodide, to afford the corresponding (tris)carbonylated compounds 8a and 8b, both as pale-yellow solids in yields of 81% and 95%, respectively. The final step of the whole synthesis was a thermal cheletropic decarbonylation of 8a/8b in solid-state form. During this process, only the planar [16]starphene 1 was formed in quantitative yield accompanied by formation of carbon monoxide as a volatile by-product.

**Characterisation of [16]starphene 1**. Following the final solid-state decarbonylation with thermal gravimetric analysis (TGA) (Figure 2 for 8a and Figure S31 for 8b) revealed that the decarbonylation, determined at 150°C, is a one-step process, through which all three carbonyls are simultaneously removed. A weight loss of 9.1% for 8a and 8.6% for 8b agrees well with the theoretical (9.2 %) weight loss of 3 carbonyls per molecule.

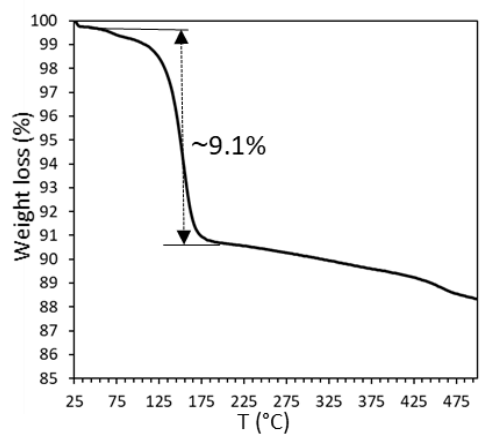

*Figure 2.* TGA of **8a** showing the weight loss of 3 CO groups (ca 9.1%; calcd 9.2 %).

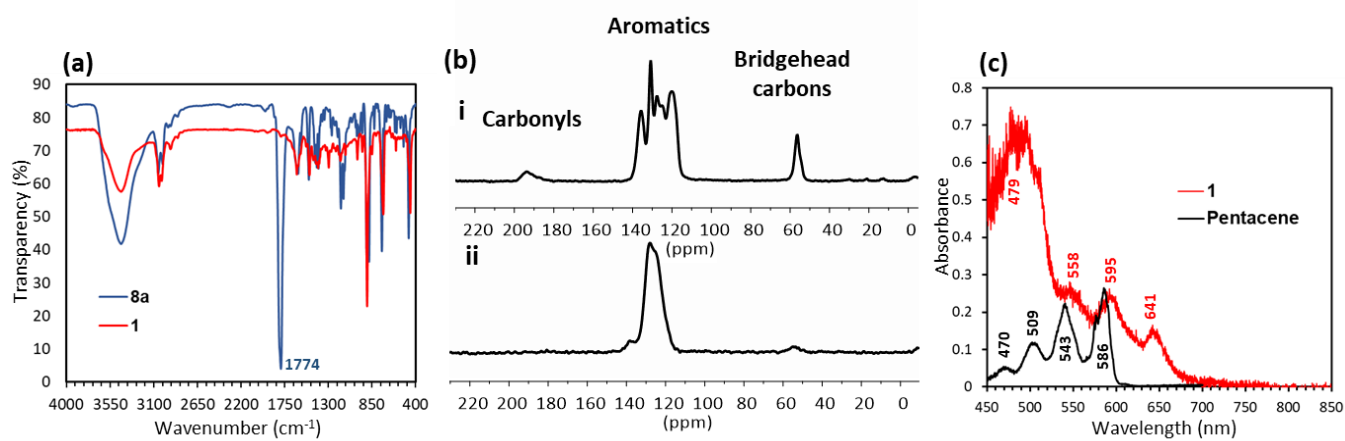

*Figure 3.* (a) FTIR spectra (KBr pellets) of **8a**; (b) $^{13}$C CPMAS NMR spectra of (i) **8a** and (ii) **1**; (c) Absorption spectra of **1** and pentacene in toluene at 5 K, concentrations $10^{-4}$ mol.l$^{-1}$.

Despite the different geometries of the isomers 8a/8b, their FTIR analyses in KBr pellets provide identical spectra and confirm the presence of carbonyl groups with a strong peak being recorded at 1774 cm$^{-1}$ (Figure 3a for 8a and Figure S25 for 8b). The isomers 8a/8b in KBr pellets were transformed to starphene 1 by heating the sample at 180 °C for 30 s. The complete decarbonylation is evident by the disappearance of a strong CO adsorption peak at 1774 cm$^{-1}$. To confirm the starphene formation, we compared the IR spectrum of 1 with that of pentacene[20] (see Supporting Information, Figure S26), which assured the presence of the pentacene moieties in the structure of 1.

To further investigate the thermal transformation of 8a/8b into 1, we recorded the carbon solid-state cross-polarisation magic angle spinning (13C CP-MAS) NMR spectra before and after decarbonylation. In the spectrum of 8a (Figure 3b, i), three groups of signals are present. The signal around 56 ppm corresponds to the bridgehead sp3 carbons, the peaks between 116 and 139 ppm belong to the aromatic carbons and a signal of carbonyl carbons is located at 194 ppm. After heating 8a at 180 °C for 15 min under an argon atmosphere, the relatively intensive signals of the bridgehead and carbonyl carbons disappeared and a narrowed signal between 114 and 140 ppm of aromatic carbons remained (Figure 3b, ii). The same behaviour was observed also for isomer 8b (Figure S21, S22). Notably, in Figure 3b, ii, a small broadened signal detectable around 55 ppm presumably belongs to the bridgehead sp3 carbons of a negligible amount of starphene dimer or higher oligomers that could be formed during the thermal decarbonylation.[25] Storing 1 in a glove box for one month at room temperature did not lead to any change in the 13C CP-MAS NMR spectrum (See Supporting Information, Figure S20) confirming its good stability in the solid-state form.

To elucidate the electronic properties of 1, we measured its absorption spectrum in a frozen matrix and compared it with that of pentacene[20,26] (Figure 3c) prepared under similar conditions from a carbonylated pentacene precursor (see Figure S29 for its structure). Solutions of 8a, 8b, respectively, were prepared in toluene with concentrations 10-4 mol.l-1 and the spectra recorded at room temperature (Figure S27a, S28a). The samples were then cooled to 5K, irradiated with UV light (310 nm) for 15 min to generate starphene 1 and measured again (see Supporting Information for further experimental details). A strong and broad absorption band with maximum at around 479 nm and then three weaker absorption bands with maxima at 558, 595 and 641 nm appear in the spectrum of 1. The obvious red shift (55 nm) of the longest wavelength absorption band of 1 compared to that of pentacene (at 586 nm) indicates electronic conjugation among the pentacene arms in 1. Based on the comparison with the literature data[26], it seems that the electronic properties of 1, diluted in frozen matrices, are similar to those of hexacene (rather than of longer acenes)Finally, MS MALDI analysis (Figure S23) of 1 revealed a mass of 828.5 Da, which agrees with the predicted value of 828.3 Da.

Theoretical calculations. The molecular orbitals of [16]starphene 1, including its Kohn–Sham HOMO and LUMO, respectively, were calculated using density functional theory (DFT) at the B3LYP/6-311++g(2d,p) level. The HOMO and LUMO are delocalised over the entire conjugated backbone of 1, and the HOMO-LUMO gap was determined as 1.64 eV, which is smaller than that of comparable calculations on pentacene (2.20 eV)[27] or hexacene (1.76 eV).[28] Together with the calculated HOMO-LUMO gap (1.95 eV) of angularly fused pentacenes (corresponding to two thirds of starphene 1), the values manifest evolution of the HOMO-LUMO gap when going from pentacene to starphene (Figure S32), thus supporting the hypothesis that the three pentacene arms are partially conjugated. To evaluate the local aromaticity of 1, nucleus-independent chemical shift (NICS) values were calculated.[29] The NICS values 1.0 Å above the molecular plane NICS(1) of 1 (Figure 1) on the benzene moieties along a pentacene arm range from –7.2 to –12.8, indicating their high degree of aromaticity. The NICS(1) values increase towards the middle rings of the arms suggesting an increase of aromaticity. On the other hand, the NICS(1) value of the central ring of 1 is +1.9, displaying almost no aromaticity.[30] This positive value suggests that single bonds interconnect the three pentacene arms and form the central ring. However, despite the positive value of the central ring, the coplanar arrangement of the three pentacene units likely allows their partial electronic coupling.

STM experiments. In order to verify the chemical and electronic structure of the [16]starphene 1 via low-temperature STM experiments, the molecules needed to be deposited onto a clean surface under ultra-high-vacuum (UHV). Attempts to sublime precursors 8a and 8b under UHV proved unsuccessful. Instead, saturated solutions of the two precursors in acetonitrile were deposited onto an Au(111) surface in high vacuum with an atomic layer injection source ("ALI", see Supporting Information for details). The samples were then post-annealed at 150°C for 5 minutes to desorb a significant proportion of the contaminants present in the solution. Large-scale STM images recorded at 4.3 K before and after the annealing process are presented in the Supporting Information (Figure S33). Very low coverage of starphene-shaped molecules was observed after this process, identified by their distinctive shape in the STM images. To our surprise, the formation of starphene-shaped structures was observed only when the precursor with a table-like structure (8b) was deposited. The isomer 8a likely desorbs before being decarbonylated. An STM image showing two of these starphene-shaped molecules is presented in Figure 4a. Functionalisation of the STM probe with a CO molecule and scanning in the repulsive CO-adsorbate interaction regime provides HR-STM images that resolve the internal structure of molecules at surfaces.[31] Such images clearly show the distinguishable differences in the middle rings of all three arms of the starphene objects found on the surface when compared to the adjacent aromatic rings (Figure 4b). While these features coincide with the carbonyl-functionalised rings of 8b, we believe a decarbonylation has already taken place and those rings are instead hydrogenated (Figure 4c). The following findings support our hypothesis: (I) occasionally, the different ring does not coincide with the originally functionalised position, but with a neighbouring ring (Supporting Information, Figure S34); (II) the post-deposition annealing treatment has been performed at a temperature at which the decarbonylation is readily known to set in (see Figure 2); (III) finally, STM images of precursors closely related to 8b, with bridging ketones, which have been used to form long acenes at surfaces in other studies, usually yield a greatly enhanced signal at the location of the ketone bridge.[32,33] In contrast to this, we typically observe a reduced signal in the centre of the arms of these objects when imaging at a voltage that is within the energy gap of the molecule (Figure S35), in line with other studies of hydrogenated acenes on surfaces.[34] Such hydrogenation breaks the conjugation along the starphene arms, resulting in what can be seen as "independent" conjugated moieties. Indeed, it has been previously shown that each of the conjugated sections hosts molecular orbitals that are extremely similar in energy and wavefunction symmetry to those of the correspondingly sized acene.[34,35] In this case, the molecule comprises naphthalene moieties at the end of each arm and a [7]starphene in the central part. In a similar fashion to hydrogenated acenes, only when imaging at energies approaching the orbitals of these remaining conjugated moieties does the darker contrast on the central ring disappear and eventually become a maximum (Figure S36)[35].

A locally different contrast on the central rings of nonacene molecules on Au(111) has also been associated with underlying Au adatoms coordinating to the corresponding C atoms, the reason being the high reactivity displayed at those positions due to the partial open-shell character of higher acenes.[32] However, such an obviously reduced signal at the different ring, as observed in our experiments (Figure S35), was not found in the mentioned nonacene case. Based on this noticeable difference in the bias-dependent contrast, we thus discard the coordination to Au adatoms as the reason for the modified appearance. We propose that, as occurs with nonacene, starphene 1 is also a very reactive molecule once it has been formed by decarbonylation of 8a and 8b, respectively.

However, under our experimental conditions, its most reactive C sites become hydrogenated during the sample processing rather than coordinating to Au. Although the source of the hydrogen atoms required for this reaction is unclear, it is presumably related to the significant presence of contaminants that were co-deposited with the molecule (Figure S33), along with the required annealing step. Note that such a pronounced reactivity as observed here for 1 has been associated with a substantial open-shell character and, in the case of polyacenes, it has only been observed with heptacene or longer molecules.[32] This can thus be readily considered a hint for a partial open-shell character of 1 and its substantially different electronic properties when compared to pentacene.

To ultimately form and characterise 1, the extra hydrogens were removed via STM tip-induced dehydrogenation. Applying bias voltage pulses of +2 V, acquiring dI/dV spectra, or even scanning at higher biases was often enough to remove the extra hydrogen atoms. An example of the sequential removal of hydrogens from each arm of a hydrogenated starphene is shown in Figure S37, with clear differences observed in STM imaging. An HR-STM image of 1 is presented in Figure 5, alongside dI/dV point spectra that display multiple resonances marked with green dashed lines. The resemblance of the conductance maps of the various resonances to the calculated wavefunctions of molecular orbitals of free-standing starphene 1 allows for their unambiguous assignment, as shown in Figure 5c for the LUMO, Figure 5d for the HOMO and Figure S38 for all the resonances together. The experimental energy gap between the HOMO (at -0.61 eV) and LUMO (at 0.9 eV) as obtained from STS is thus 1.51 eV, which is in a good agreement with the calculated value of 1.64 eV (vide supra) and is comparable to the gap measured with STS for heptacene on gold.[36],[37].

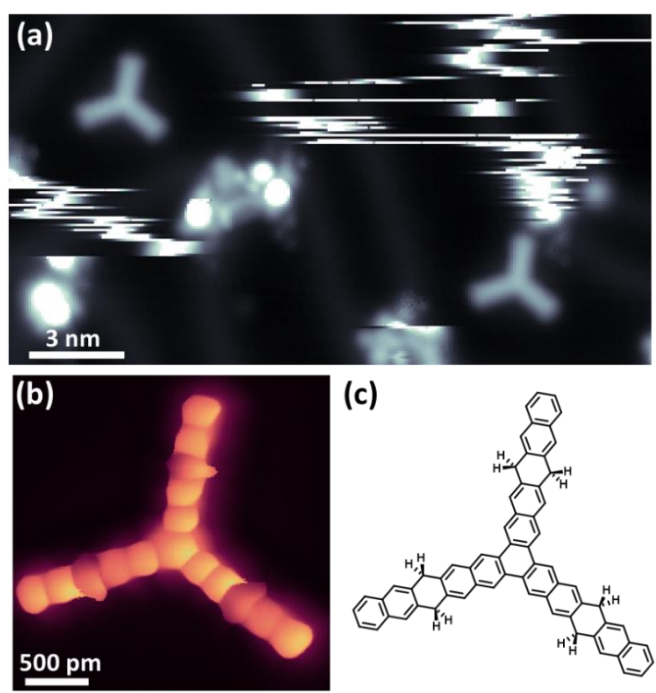

*Figure 4.* (a) STM image (−600 mV, 50 pA) of the starphene/Au(111) sample after deposition and post-annealing to 150°C. Some of the objects on the surface were often moved around by the tip, leading to bright streaks in the image. (b) HR-STM image (CO tip, constant height, 2 mV) of the hydrogenated starphene prior to hydrogen removal via the STM tip. (c) The expected corresponding chemical structure of the hydrogenated starphene. All STM images recorded at 4.3 K.

The HOMO-LUMO gap reduction by about 30% with respect to that of pentacene on Au(111) (2.20 eV as measured by STS)[38] provides further evidence that the molecule does not entirely behave as three isolated pentacene arms, but instead shows a notable electronic coupling between them.

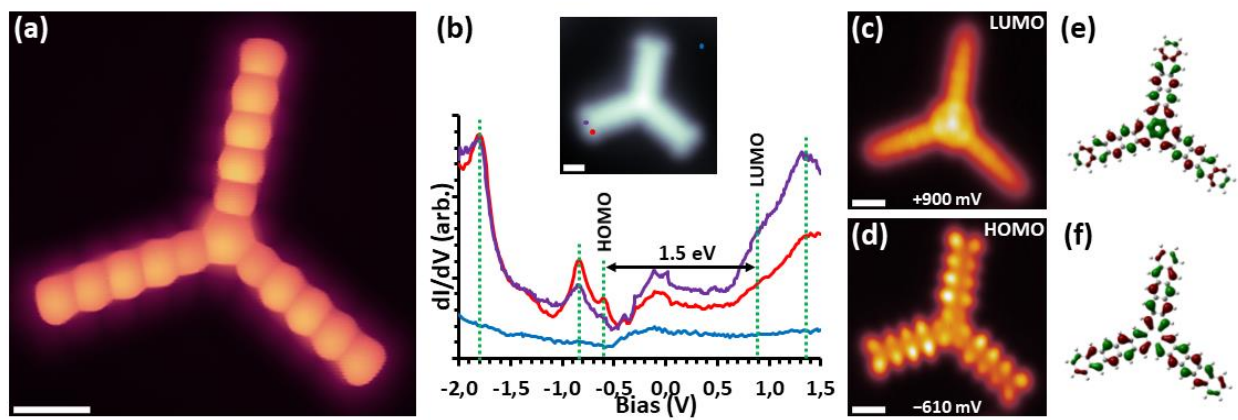

*Figure 5.* (a) HR-STM image (CO tip, 5 mV, constant height) of a starphene after hydrogen removal. (b) dI/dV point spectra on starphene showing various resonances marked with dashed lines. The inset marks the positions at which each of the spectra are taken, two on the molecule and one on the substrate as reference. (c) and (d) Constant height dI/dV images of a starphene recorded with a Cl tip at the energy of the lowest

**Conclusion**

We have demonstrated an efficient new strategy towards the preparation of unsubstituted [16]starphene 1. We believe that our general strategy is also adaptable for the construction of substituted analogues and, potentially, for even larger starphenes than 1 presented here. The multi milligram scale synthesis of 1 enabled its comprehensive characterisation by a variety of analytical methods, such as 13C CP-MAS NMR, TGA, MS MALDI and FTIR spectroscopy. UV-Vis absorption spectrum of 1 in frozen toluene was red-shifted compared to that of pentacene, indicating a partial electron conjugation of the starphene arms. The HR-STM experiments provided a characterisation of the fundamental structural and electronic properties and confirmed the moderate decrease of the HOMO-LUMO gap with respect to pentacene.

Our synthetic strategy enables to produce starphene compounds in large quantity, which is essential for the investigation of electron/hole mobilities in bulk and offers the possibility to implement 1 into electronic devices such as OFETs and OLEDs.

**Acknowledgements**

This research has been funded in parts by foundation EXPERIENTIA, ERDF/ESF "UOCHB MSCA Mobility" (No. CZ.02.2.69/0.0/0.0/17 050/0008490), and has received funding from the EraNET Cofund Initiatives Quan-tERA under the European Union's Horizon 2020 research and innovation programme grant agreement ORQUID. This project has received financial support from the CNRS through the MITI interdisciplinary programs and JST-ERATO (No. JPMJER1903) and JSPS-WPI. Financial support is also acknowledged from the Spanish Ministry of Science and Innovation

(MAT2016-78293-C6, PID2019-107338RB-C63), from the Basque Government (IT-1255-19), from the Spanish National Research Council (CSIC, COOPB20432), from the Euro-pean Regional Development Fund (ERDF) under the program Interreg V-A España-Francia-Andorra (contract no. EFA 194/16 TNSI) and from the European Research Council (grant agreement no. 635919). The authors thank Yannick Coppel (LCC-CNRS, Toulouse) for recording CP-MAS spectra, Jean-Francois Meunier (LCC-CNRS, Toulouse) for thermogravimetric analysis and Vladimir Vrkoslav (IOCB-Prague) for recording MS MALDI spectra.

**Keywords**: [16]Starphene, solid-state synthesis, decarbonylation, scanning tunnelling microscopy, π-conjugation, HOMO-LUMO gap, acenes.